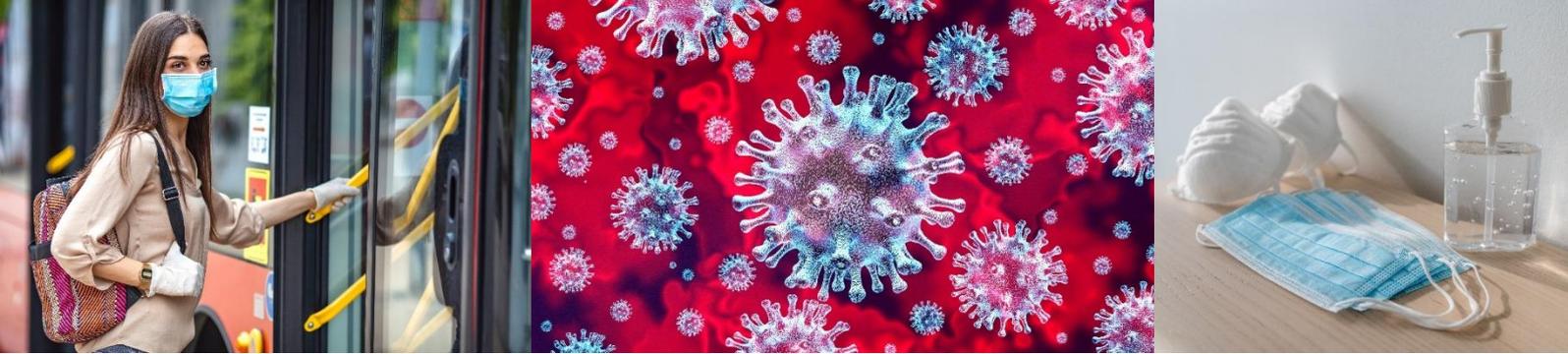

# IMPACT OF COVID-19 ON TRAVEL BEHAVIOUR, TRANSPORT, LIFESTYLES AND RESIDENTIAL LOCATION CHOICES IN SCOTLAND

Authors: Lucy Downey, Achille Fonzone, Grigorios Fountas and Torran Semple

Transport Research Institute, Edinburgh Napier University, Colinton Road, Edinburgh, EH10 4DT
https://blogs.napier.ac.uk/tri/projects/

Funded by a grant from the Scottish Funding Council

**Abstract**

COVID-19 was declared a pandemic by the World Health Organisation (WHO) on 21$^{st}$ March 2020 and on 24$^{th}$ March 2020, the UK and Scottish Governments imposed a 'lockdown' restricting everyday life activities to only the most essential. These Governmental measures together with individual choices to refrain from traveling during the COVID-19 pandemic have had a profound effect on transport related activity.

In the current investigation an online questionnaire was distributed to 994 Scottish residents in order to identify travel habits, attitudes and preferences during the different phases of the pandemic outbreak and anticipated travel habits after the pandemic. Quota constraints were enforced for age, gender and household income to ensure the sample was representative of the Scottish population as a whole.

Perceptions of risk, trust in information sources and compliance with COVID-19 regulations were determined together with changes in levels of 'life satisfaction' and modal choice following the onset of COVID-19. In addition, survey responses were used to identify anticipated travel mode use in the future. Consideration was also given to the effects of COVID-19 on transport related lifestyle issues such as 'working from home', online shopping and the expectations of moving residences in the future. As part of the analysis, statistical models were developed to provide an insight into both the relationships between the levels of non-compliance with COVID-19 regulations and demographic variables and the respondent attributes which might affect future public transport usage.

In general, the study confirmed significant reductions in traffic activity, amongst respondents during the COVID 19 pandemic associated with walking, driving a car and either using a bus or train. The respondents also indicated that they anticipated they would continue to make less use of buses and trains at the end of the pandemic.

# TABLE OF CONTENTS

# LIST OF TABLES

# LIST OF FIGURES

# 1. INTRODUCTION AND BACKGROUND

## 1.1. Introduction

COVID-19 was declared a pandemic by the World Health Organisation (WHO) in March 2020. The UK and the Scottish Governments initially responded to the threat of COVID-19 by imposing a 'lockdown' to restrict everyday life activities to the essential minimum. Individuals choosing to limit their travel, in order to reduce their own exposure to COVID 19, together with the effects the Government imposed lockdown significantly affected transport and travel patterns. There was, for example, a dramatic decrease in vehicular activity and many more employees started working from home.

## 1.2. Transport and Travel in Scotland

Table 1.1 summarises key government measures to contain the virus spread through restricting travel and transport in Scotland. Since the onset of the COVID-19 pandemic, the UK Government and the devolved administrations[1] have had to enforce massive restrictions on public transport in order to limit transmission of the virus and ensure safe passage of key workers during the emergency response.

**Table 1. 1 Summary timeline of key events and COVID-19 restrictions in Scotland**

| Date | Event | Description |
|---|---|---|
| 11th March 2020 | WHO declares COVID-19 a pandemic | On 31 December 2019, Chinese authorities notified the World Health Organisation (WHO) of an outbreak of a new disease in Wuhan. The subsequent spread of the virus to other countries led to WHO declaring the outbreak a public health emergency. |
| 24th March 2020 | UK National Lockdown | The UK Government enacted emergency legislation to restrict personal movement. These 'lockdown' restrictions limited travel and transport outside the home to essential journeys and exercise. Public transport operated with limited service and capacity |
| 29th May 2020 | Scottish Governments Route map out of lockdown Phase 1 | Easing of lockdown restrictions allowing people to meet family and friends outdoors. However, physical distancing and staying at home still key. |
| 18th June 2020 | Scottish Governments Route map out of lockdown Phase 2 | Includes meeting more people and shops reopening; public transport services increase over the phase, but capacity remains constrained due to physical distancing requirements. Active travel remains the preferred mode of travel |
| 10th July 2020 | Scottish Governments Route map out of lockdown Phase 3 | Different households can meet indoors<br>Can drive beyond local area for leisure and exercise purposes.<br>Public transport continued to scale up to full services during this phase. However services operated at reduced capacity due to physical distancing requirements |

---

[1] UK devolution refers to the distinct legislatures and governments in Scotland, Wales and Northern Ireland which have power over a range of policy areas which had previously been the preserve of the UK Government

| Date | Event | Description |
|---|---|---|
| 2nd November 2020 | Scottish Governments five-level strategic framework comes into force. | COVID protection levels set out measures that can be applied nationally or locally depending on the prevalence of the virus across Scotland. |
| 5th January to present (12th February 2021) | UK National Lockdown | Mainland Scotland goes into lockdown with new legal requirement forbidding anyone from leaving their home except for essential purposes. |

*Sources: Scottish Government (2021); Scottish Parliament (2021)*

A Scottish survey (Transport Scotland 2021a) indicated that concerns about using public transport remain high with 69% of those surveyed being fairly or very concerned about contracting or spreading the virus while using public transport and 62% of those surveyed fairly or very concerned about having enough space to observe physical distancing on public transport. Almost half of respondents (46%) stated that, in the future, they would avoid public transport and use their car more than they did before the pandemic. Similar findings were obtained from UK Transport Focus Survey (2021). A third of respondents said that even when COVID-19 no longer poses a significant risk they would continue to drive more and use public transport less.

## 1.3. Current Research

In response to the COVID-19 pandemic, Edinburgh Napier University's Transport Research Institute has been undertaking a study, funded by the Scottish Funding Council (SFC), into its impact on transport and travel in Scotland. As part of this research, a travel behaviour questionnaire was developed focusing on daily travel as well as people's long-term travel habits, attitudes and preferences during the different phases of the pandemic outbreak. An insight into how COVID-19 might change travel choices in the near future and beyond has the potential to assist both the UK and Scottish Governments and transport providers in ensuring both the future resilience and the efficiency of the transport system, particularly when future events may result in major mobility restrictions. It was hoped that this investigation would yield information to assist in developing appropriate COVID-19 transport related responses, including policies, interventions and communications in addition to the messaging necessary to encourage uptake of protective measures.

In addition, it was hoped that the research would contribute to an understanding of the potential long-term consequences of transport and travel related behaviour such as the desire to change residential location post COVID-19, the likelihood of households purchasing additional vehicles and likelihood of sustainable travel choices using public transport and/or active travel modes. Information on residential choices will assist in predicting future public transport demand, especially in Scotland.

The first part of this report describes the methodology developed to undertake the online survey. The remainder of the report provides details of the results and analysis, under the headings listed below followed by the associated discussion and conclusions:

- Communication of COVID-19 Information and Responses (sources of information, trustworthiness of sources, compliance with COVID-19 transport regulations and guidance and overall perceptions of the COVID-19 pandemic);
- Travel Choices, Attitudes and Perceptions (perceived risk of travel by mode, perception of effectiveness of mitigation measures and travel by mode before, during and potentially after the pandemic); and
- Lifestyle Issues (working from home, online shopping and potential relocation of residences).

## 2. METHODOLOGY

### 2.1. Online Survey

In order to establish an understanding of perceptions and attitudes of residents in Scotland as well as their responses to the COVID-19 pandemic, a survey was conducted using the online platform, Qualtrics. The associated questionnaires were completed by participants between 3rd February 2021 and 17th February 2021. The survey was restricted to Scottish residents and involved enforcing quota constraints for age, gender and household income to ensure the sample was representative of the Scottish population.

A managed pilot study was initially conducted on 2nd February 2021 to review the setup, check data quality, questionnaire performance, incidences of potential straight lining, speeding and invalid responses. The median time for survey completion was 17 minutes and the mean time was 21 minutes. A speeding check (measured as half the median pilot survey time) was added which automatically terminated those who were not responding thoughtfully.

In order for us to comply with UK General Data Protection Regulation (UK GDPR), potential survey participants were provided with a privacy notice detailing what we will do with the survey data. Informed consent was required before the respondent could participate in the study.

The questionnaire was structured into a series of short, largely closed-ended items covering the following topics:

- Perceptions of COVID-19 (impact on life; risk perceptions associated with using different types of transport modes, life satisfaction);
- Information and advice about COVID (frequency of seeking advice, trustworthiness of different source, compliance with government guidance and regulations);
- Mode choice (before COVID, during various stages of lockdown and anticipated mode choice in the future);
- Travel related activities (e.g. working from home, online shopping etc.);
- Perceived effectiveness of various measures to reduce the spread of COVID-19 on public transport;
- Residential location choices (likelihood of relocating, reasons for moving); and
- Socio-demographic characteristics (age, gender, ethnicity, education qualifications, employment status, household income, car availability, household type and residential location).

## 2.2. Analysis

A total of 994 responses were collected. Missing and 'don't know' responses were not included in the analysis unless otherwise indicated. Where the latter was the case reference is made to this action in the notes attached to each table. Where numbers have been rounded up, in the analysis, the sum of the constituent items may not necessarily match the actual totals shown.

The statistical package SPSS was utilised to analyse and compare responses. Statistical comparisons were typically accomplished via a Chi-Squared test, a non-parametric test for identifying statistically significant differences between groups of observations defined by one or more categorical variables. The remainder of this report describes survey data from the online survey.

# 3. PROFILE OF RESPONDENTS

In this Section consideration is given to how representative the survey sample was to that of those living in Scotland as a whole both in terms of socio-demographic characteristics and in terms of transport modes utilised by respondents prior to the COVID-19 pandemic.

## 3.1. Gender, Age and Other Respondent Details

Figure 3.1, Figure 3.2, Figure 3.3 and Figure 3.4 provide details covering the gender split, age distributions, income distributions and the regional distributions of the survey respondents' residences, respectively. The Figures also provide details of the equivalent distribution, for Scotland as a whole, which were derived from Census data obtained from National Records of Scotland (2019 mid-year population estimates). Figure 3.2 shows that the survey responses involved a smaller proportion of those aged 45 or older than those for the Scottish population as a whole. Furthermore, the proportion of high-income households (greater than 15k) was higher than that of the Scottish population (as shown in Figure 3.3). Previous research has indicated that some demographic variables such as age and income affect participants' willingness (or opportunity) to complete online surveys. For example, Millar et al (2009) found that respondents to internet surveys were younger than mail survey respondents, had higher levels of education and higher incomes. The use of the internet is not homogeneous across all population groups and further research should examine the existence of possible bias.

In addition to determining age, gender, income and region, the survey identified ethnicity, educational attainment, employment status and household car ownership. These characteristics are summarised as follows:

- 96% (n=993) of respondents would describe their ethnic group as 'white';

- Over half (56%) of respondents were employed (either part-time of full-time) and 21% were retired at the time of data collection. The remaining 23% were either in education, looking after home, full time carers or unemployed; and

- Over a fifth (21%) of respondents lived in households without access to a car and 29% lived households with two or more cars available.

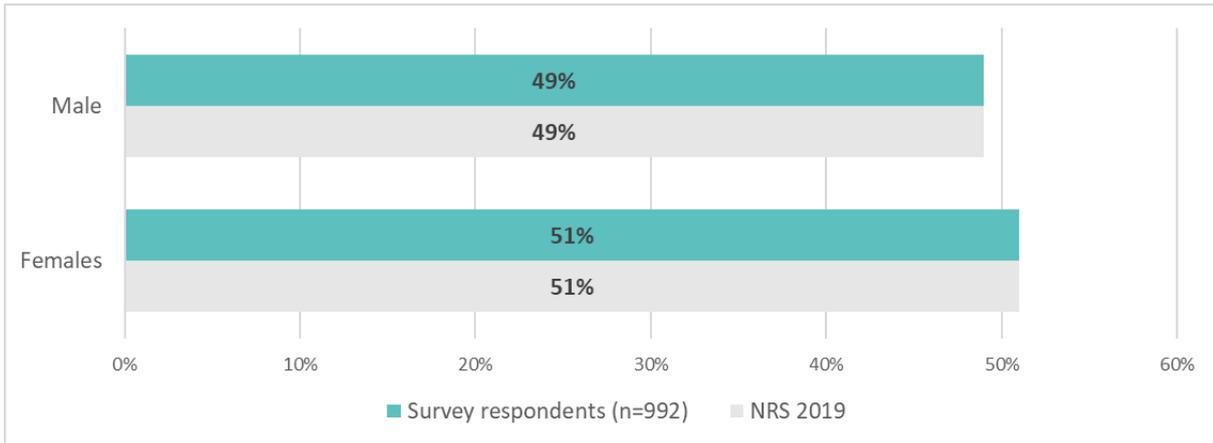

**Figure 3. 1 Gender distribution of survey respondents and Scottish population (NRS)**

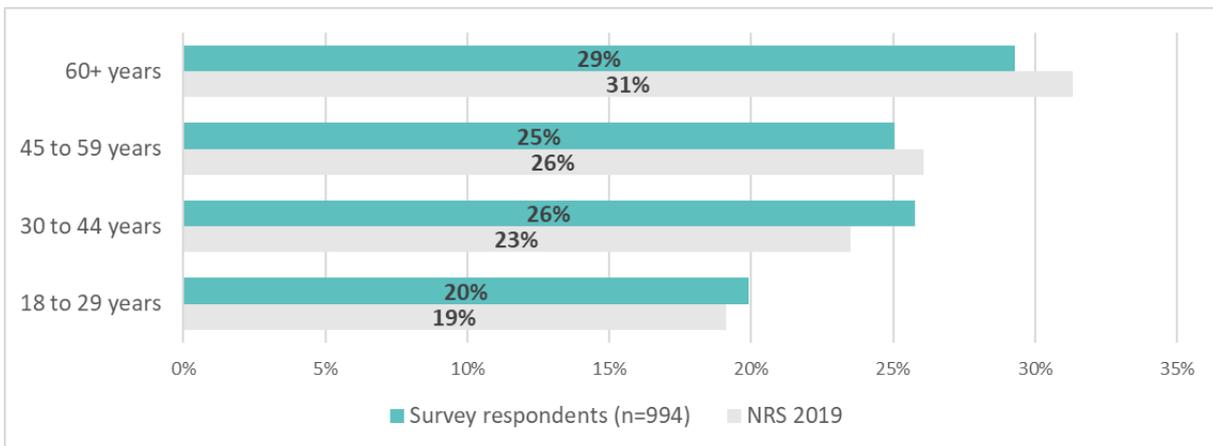

**Figure 3. 2 Age distribution of survey respondents and Scottish population (NRS)**

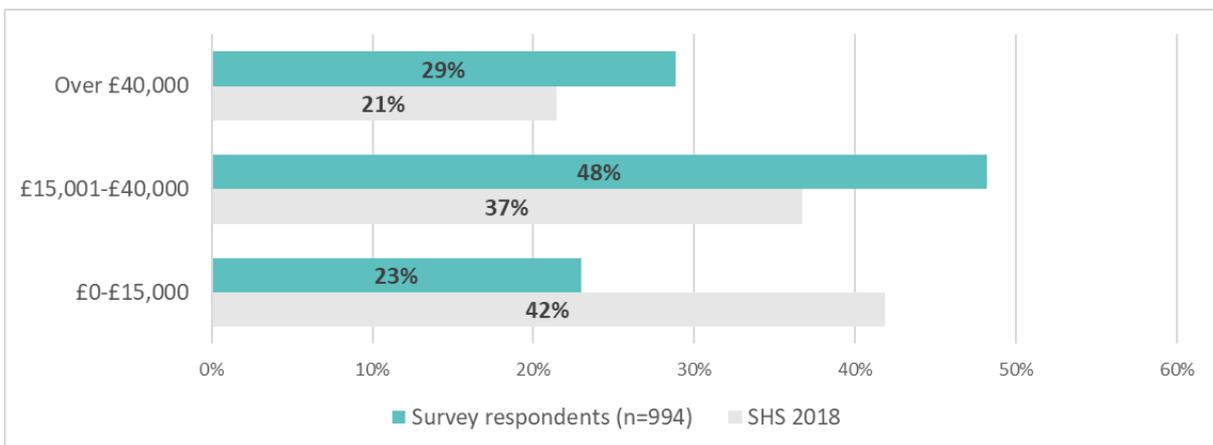

**Figure 3. 3 Income distribution of survey respondents and Scottish population (SHS)**

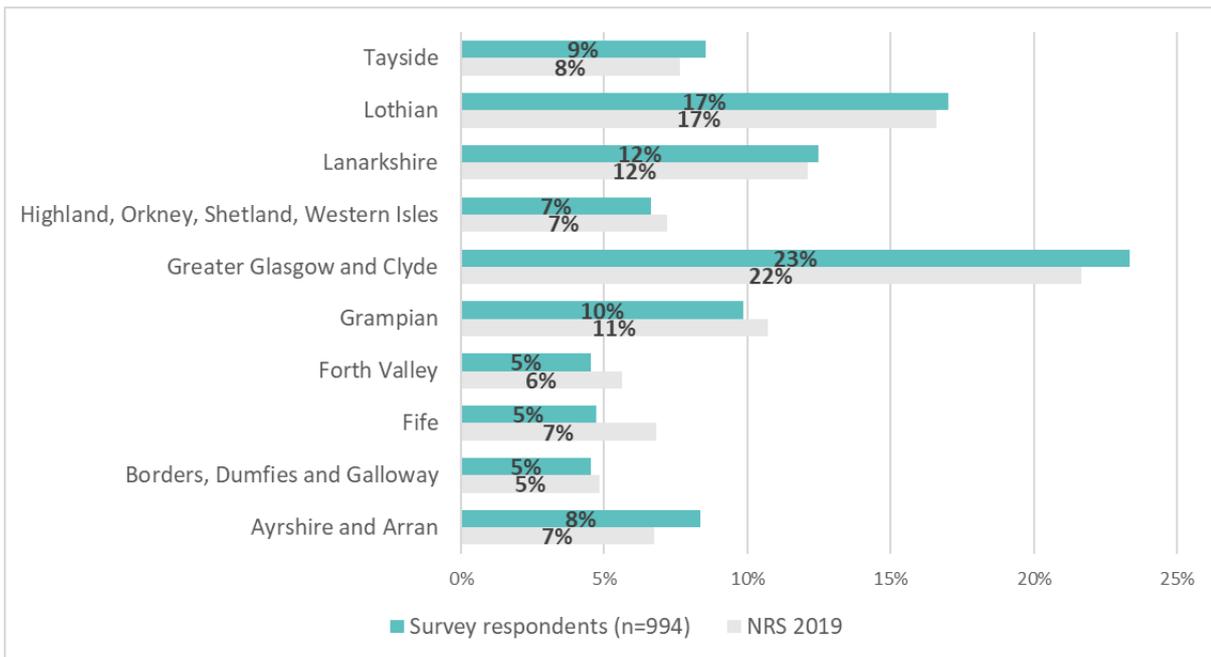

**Figure 3. 4 Residential location distribution of Survey respondents & Scottish population**

## 3.2. Mode Choice Prior to Lockdown

Figure 3.5 shows the frequency with which the respondents stated they used cars (as driver and passenger), public transport, cycling and walking before the COVID-19 pandemic (i.e., before 11[th] March 2020). It may be seen, from Figure 3.5, that prior to the COVID-19 pandemic, that on at least three days a week 60% of respondents walked, 57% drove a household vehicle, 19% travelled as passengers in a household vehicle, 16% travelled on a bus, 7% travelled by train and 6% rode a bicycle.

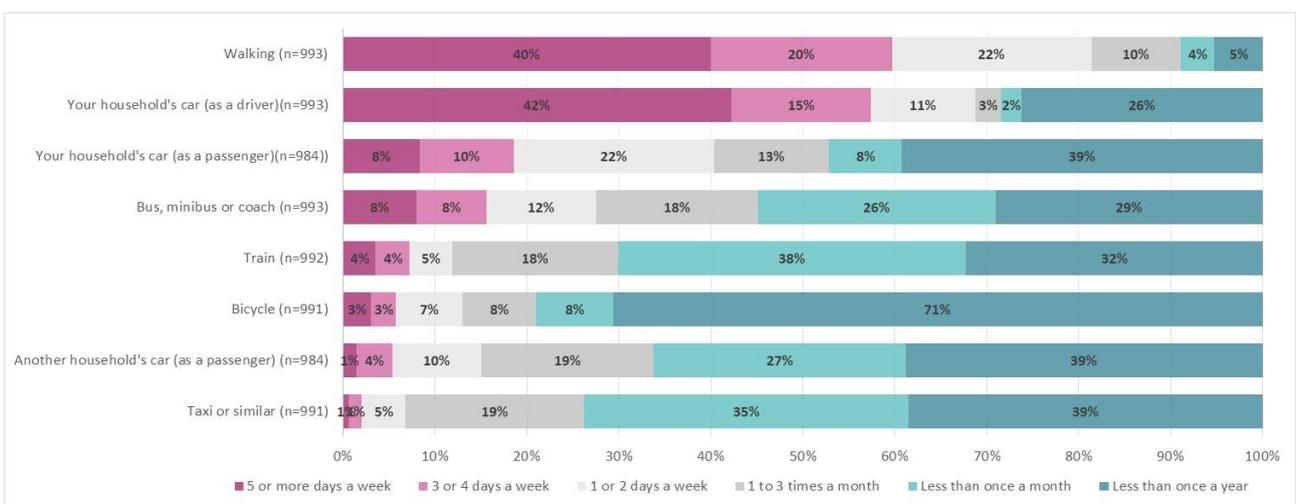

**Figure 3. 5 Frequency of use of different forms of transport**

## 4. RESULTS AND ANALYSIS

### 4.1. COVID-19 and Life Satisfaction

The current study found that, before the World Health Organisation (WHO) declared COVID-19 a global pandemic almost three-quarters (72%) of those responding, indicated that they were either 'slightly satisfied' or 'very satisfied' with the way things were going in their life. However, in February 2021, during the lockdown in place at the time of the survey (i.e. lockdown 2), 66% of respondents indicated that life had got worse since then.

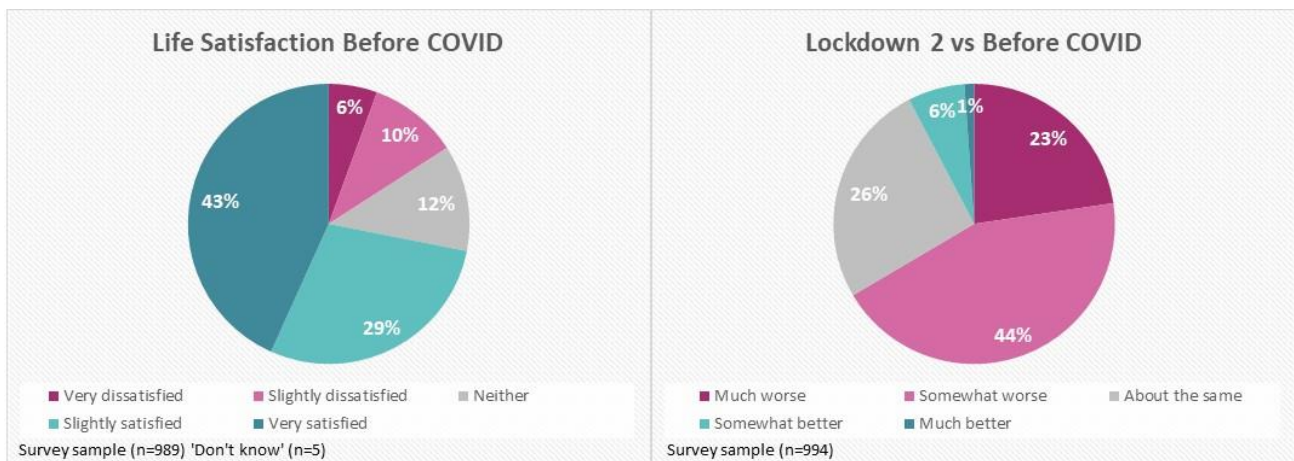

**Figure 4. 1 Life satisfaction before COVID-19 and during lockdown**

### 4.2. Communication of COVID-19 Information and Responses

#### 4.2.1. COVID-19 information sources and trustworthiness

The precautionary approach to the Covid-19 pandemic depended on effective risk communication. Consequently, it is important to understand the different COVID-19 information channels people choose, together with their level of trust in them.

Study participants were asked how often they used different channels of information to stay informed about COVID-19 and the trustworthiness of information sources. It may be seen from Figure 4.1 that 70% of the sample 'often' or 'always' used tradition broadcasters (radio and television) for information. In contrast, only 27% regularly used newspapers as a source of COVID-19 information.

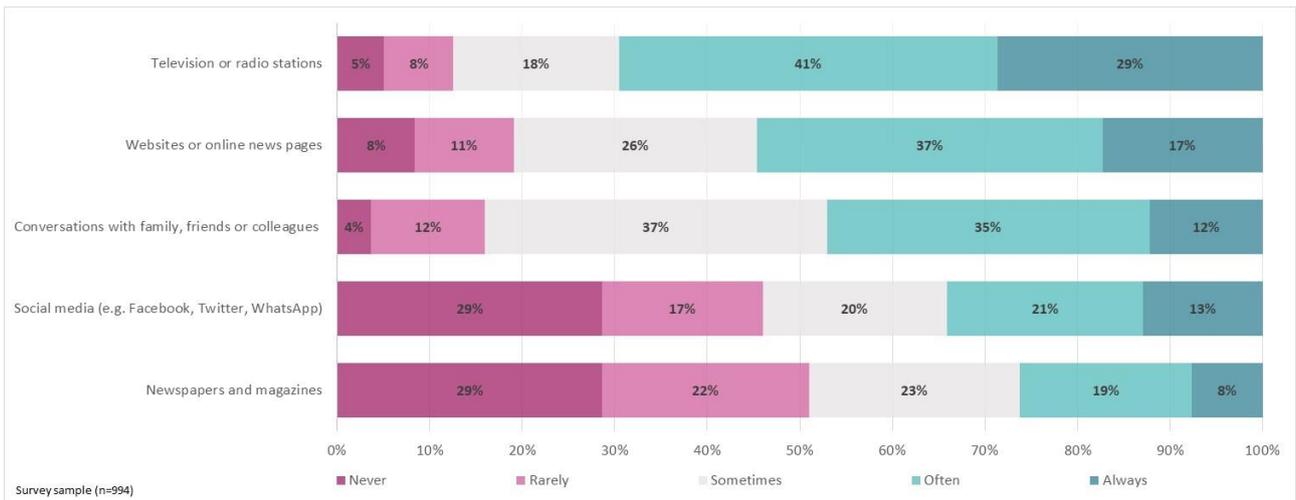

**Figure 4. 2 COVID-19 information sources**

In terms of trustworthiness of the sources of COVID-19 information, most respondents trusted medical experts (80% of respondents selected either 'extremely' or 'very' trustworthy) and the World Health Organisation (67% of respondents selected either 'extremely' or 'very' trustworthy). Just over half (51%) of the respondents considered the Scottish Government to be a very trustworthy source of information. In contrast, only 26% considered the UK Government to be a very trustworthy source.

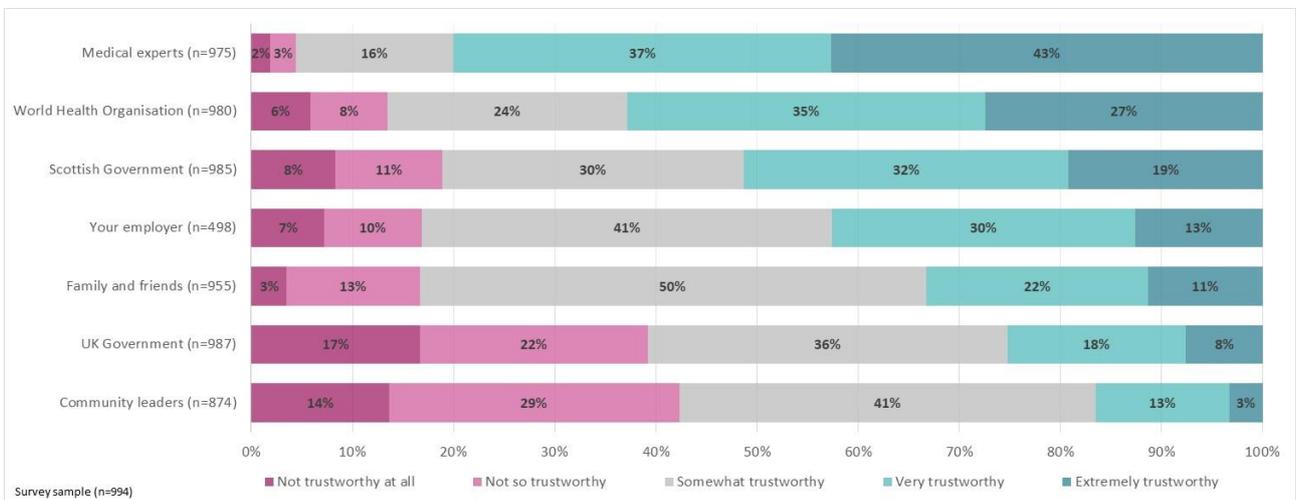

**Figure 4. 3 Trustworthiness of information and advice about COVID-19**

### 4.2.2. Compliance with Government guidelines

Respondents were asked to what extent they are following the recommendations from Scottish Government COVID-19 regulations and guidance on travel and transport, ranging from 1 (not at all) to 7 (completely). It is worth noting that we ask participants to self-report their compliance, which relies on participants understanding the regulations. Figure 4.3 shows that complete compliance with the rules (i.e., following them with no bending or even minor infringements) was reported by over half (57%) of the respondents. Most respondents (83%) reported following the recommendation to a large extent (with a score of 6-7). As may be seen from Figure 4.3, this confirms Transport Scotland findings (February 2021) that the vast majority (89%) state that they are following the regulations and guidance on travel and transport completely or nearly completely.

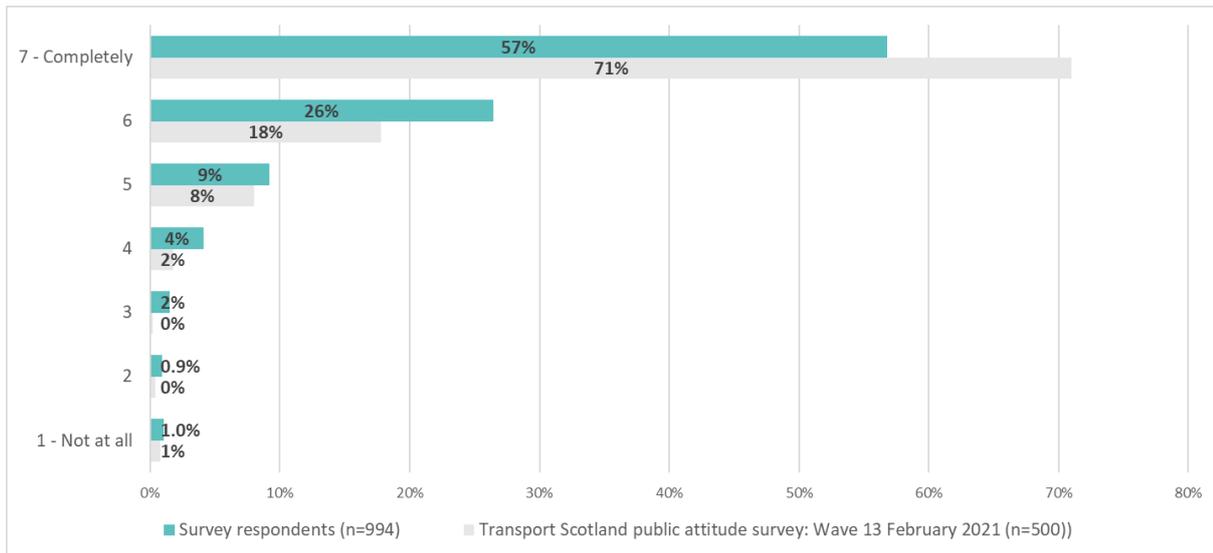

**Figure 4. 4 Compliance with COVID-19 Government transport and travel restrictions survey respondents and Transport Scotland (2021) public attitude survey**

Consideration was given to factors that significantly influenced levels of compliance among respondents. Some of the most important demographic influences were as follows:

- younger respondents (under 25) were significantly more likely than other age groups to not comply at all, while elderly individuals (over 65) were significantly more likely to comply with all regulations;
- females were found to be significantly more likely than other genders to comply with regulations fully; and
- those who were unable to work because of long-term illnesses or disabilities were also more likely to comply fully with COVID-19 travel regulations.

In addition to demographic influences, several attitudinal traits of respondents, mostly related to trust in governments or information sources for COVID-19 related guidance, were found to significantly influence compliance:

- Those who consider the Scottish Government to be a very trustworthy source, with regards to information related to COVID-19, were significantly more likely to comply fully when compared with those who perceived the Scottish Government as untrustworthy.
- Additionally, those who frequently used television or radio to stay informed about COVID-19 were significantly more likely (than those who don't use this information source) to comply fully. Those who frequently use websites or online sources were also significantly more likely to comply fully.

### 4.2.3. Overall perceptions of the COVID-19 pandemic

Respondents were asked to rate their level of agreement with statements relating to their concerns about the COVID-19 pandemic and the impact it has had on their lives. With reference to Figure 4.5 it may be seen that 48% of the respondents felt that COVID-19 was a danger to them (rated either '1' or '2' on a scale of 1 to 5). The majority of respondents (86% rated either '1' or '2') felt they were well informed about COVID-19. Regarding the impact of COVID-19, over two-thirds (69% rated either '1' or '2') thought that COVID-19 had completely changed their lives.

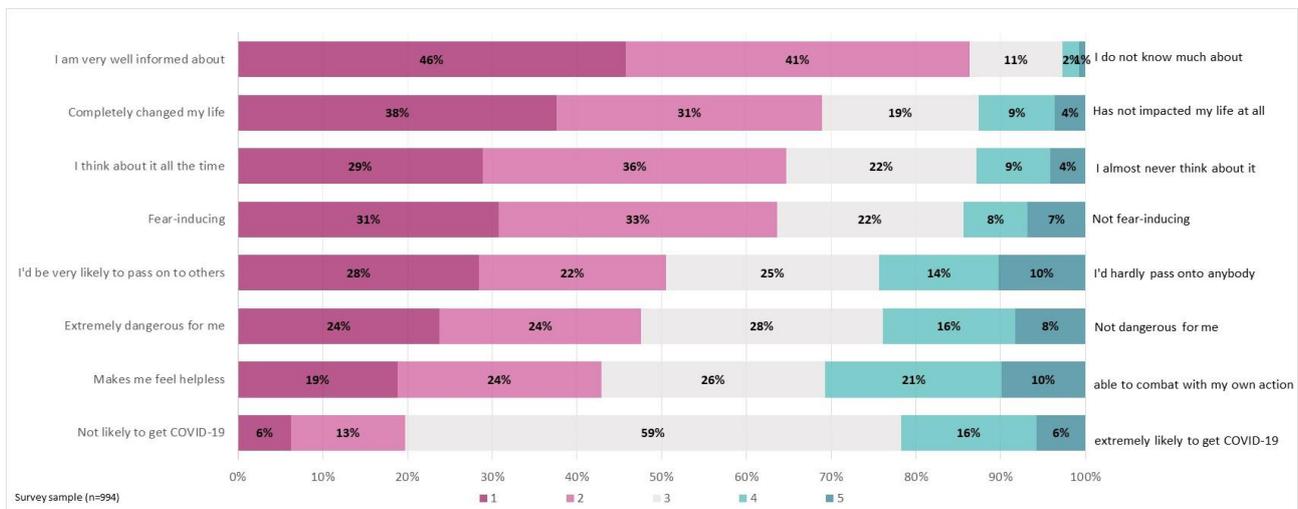

**Figure 4. 5 Perceptions of COVID-19**

## 4.3. Travel Choices, Attitudes and Perceptions

### 4.3.1. Perceived risk of travel by mode

Respondents were asked to rate how risky different types of transport would be in terms of contracting or spreading the virus. As may be expected, the majority of respondent viewed walking (92%) and cycling (93%) as low risk. In contrast, with reference to travelling by bus or train, 64% and 57% or respondents respectively considered these to be high-risk activities. Travelling by aeroplane was considered the most dangerous mode with 84% of respondents considering it was a high-risk activity.

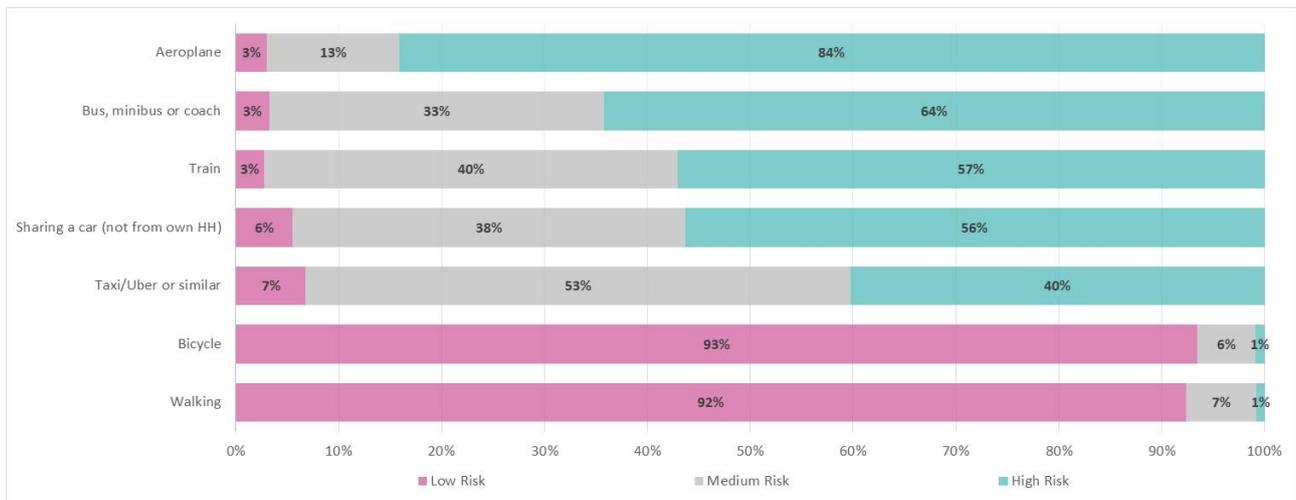

**Figure 4. 6 Perception of risk and type of transport**

### 4.3.2. Perception of effectiveness of mitigation measures

The results indicate that in general all mitigation measures stated in the questionnaire were largely considered to be important to respondents in terms of the potential to reduce the spread of COVID-19 on public transport. The preventative measures perceived to be most important were hand hygiene (93% rated either 'very important' or 'extremely important') and cleaning of station, stops and vehicles (93%). Social distancing between people at stops (92%) and on-board public transport (93%) were also viewed as 'very important' or 'extremely important'.

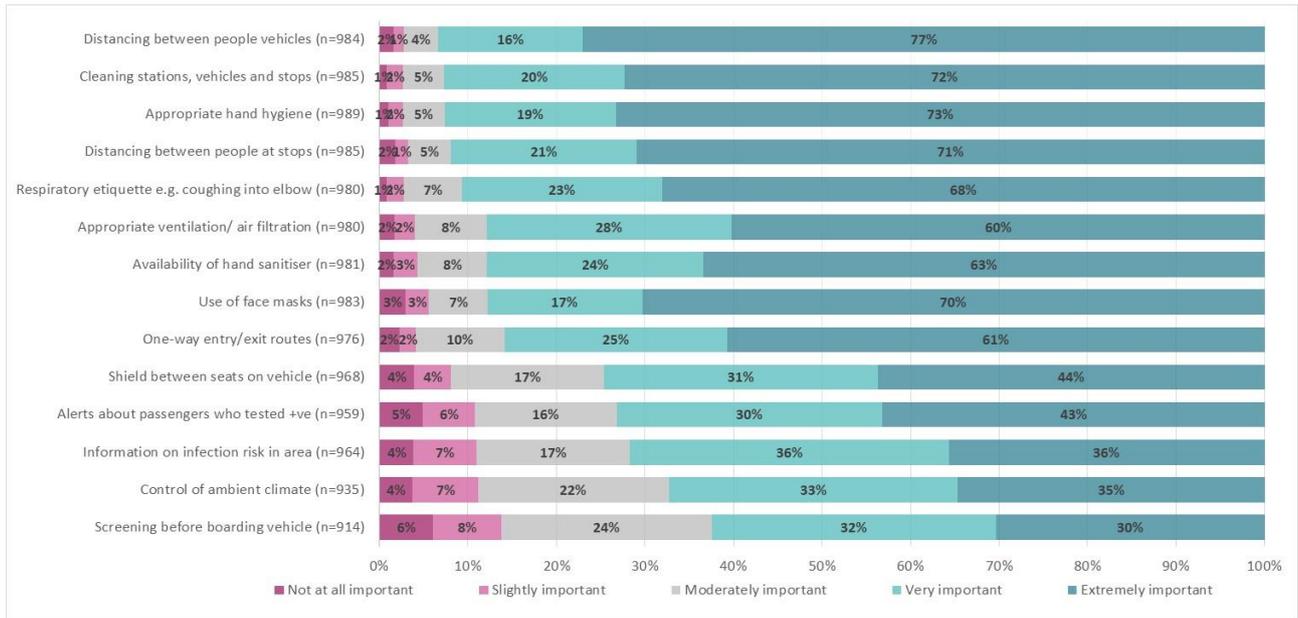

**Figure 4. 7 Public transport measures to reduce the spread of COVID-19**

### 4.3.3. Transport activity before and after COVID-19

Respondents were asked how frequently they used cars (as driver and passenger), public transport, cycling and walked more than half a mile before and during the second COVID-19 pandemic lockdown (lockdown 2 in place since 5[th] January 2021). There was a decrease in those walking at least half a mile at least once a week from 81% to 75%. The percentages remained the same for cycling and those driving their own cars reduced from 69% to 55%. For those using public transport at least once a week, the number using buses fell from 27% to 9% and the percentages using trains fell from 12% to 4%.

Respondents were asked if they currently (i.e., during lockdown 2) travelled 'less often' or 'more often' than during lockdown 1 (24th March 2020 to 29th May 2020). It may be seen, from Figure 4.9, that 43% of respondents reported using buses less often when compared with lockdown 1. The same percentage (43%) reported using trains less often.

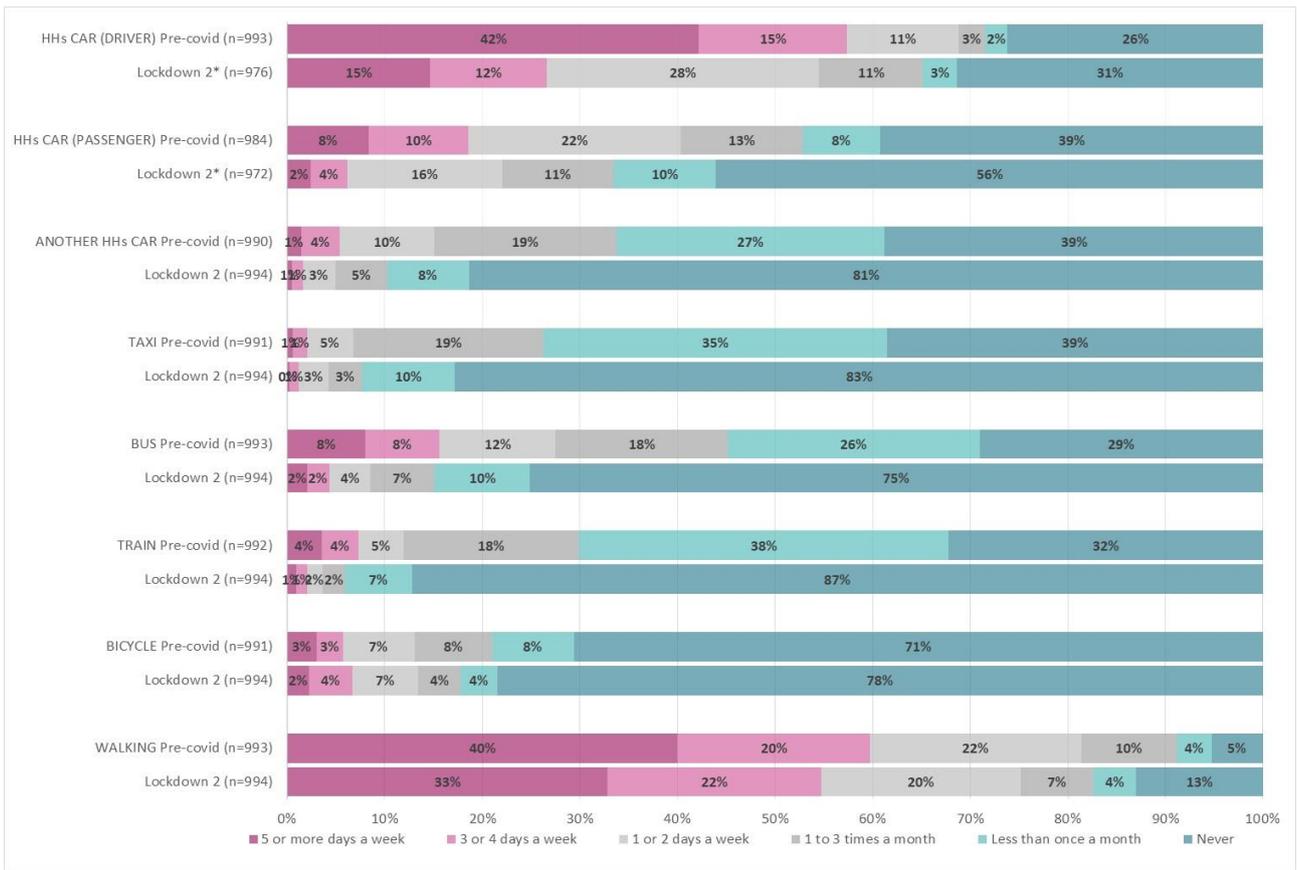

**Figure 4. 8 Travel mode pre-COVID and during lockdown 2**

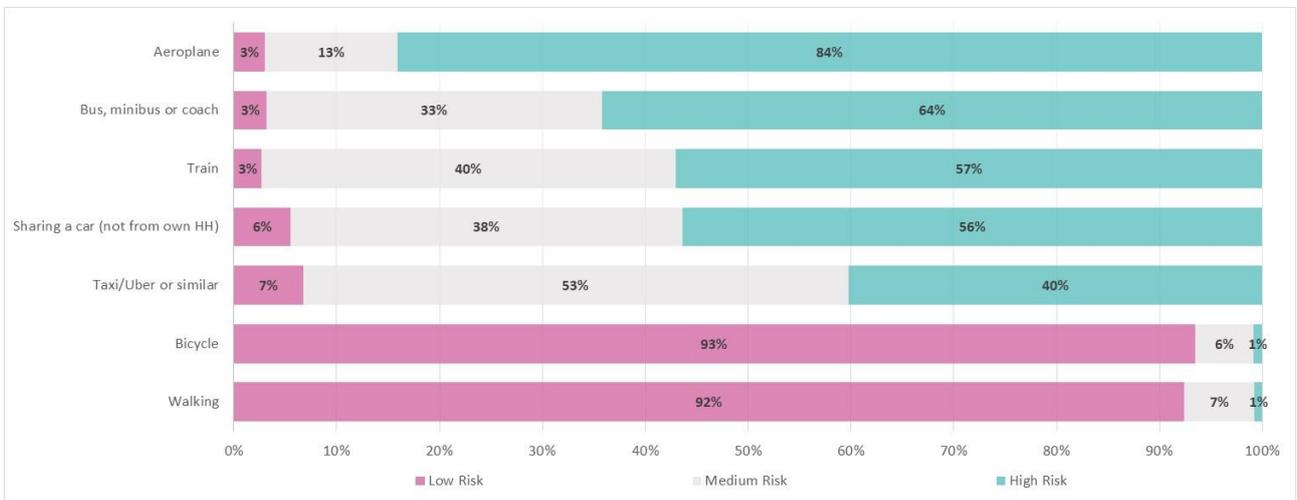

**Figure 4. 9 Travel during lockdown 2 compared to travel during lockdown 1**

### 4.3.4. Future transport modes

Survey responses indicated that 45% or respondents expected to walk more, 29% expected to cycle more and 25% expected to drive their car more post COVID-19 (in 12 to 18 months ahead) assuming an effective vaccine has been deployed on a large scale. In contrast, 42% of respondents anticipated using aeroplanes less, 36% using buses less and 34% using trains less. It should be noted that the sample size is low for some modes due to respondents selecting either 'do not know' or 'not applicable'.

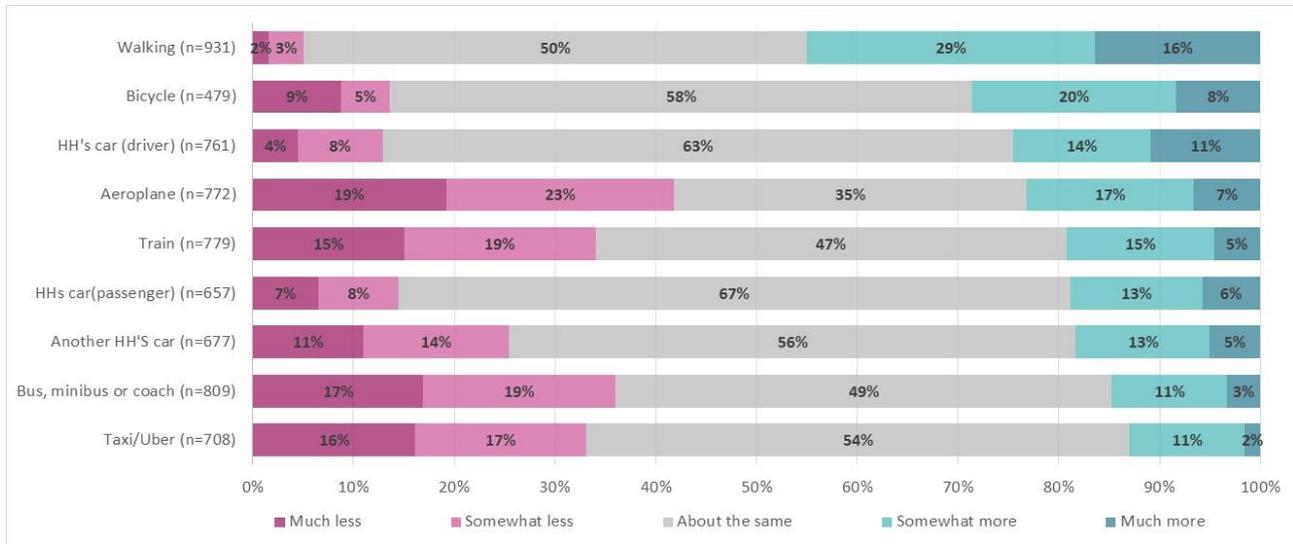

**Figure 4. 10 Anticipated future mode usage**

**Table 7. 1 Reasons for using public transport less in the future**

| Reason | Number of Responses* | Percentage* (n=333) |
|---|---|---|
| Possibility of getting infections (e.g. COVID-19) carried by other passengers | 210 | 63% |
| Lack of cleanliness/hygiene on board public transport | 163 | 49% |
| Public transport is too crowded | 151 | 45% |
| I do not like travelling with strangers | 72 | 22% |
| Public transport is too slow and/or takes too long | 71 | 21% |
| Public transport is too expensive | 64 | 19% |
| Public transport is unreliable | 60 | 18% |
| Public transport service not regular enough (infrequent) | 50 | 15% |
| Public transport is too polluting | 49 | 15% |
| Public transport is not available for my usual trips | 47 | 14% |
| Public transport service starts too late or finish too early | 36 | 11% |
| Health condition - difficult walking to/from the stop, get on and/or off the vehicle | 27 | 8% |
| Working from home and/or use video conferencing | 13 | 4% |
| Use other mode (car, cycle, walk) | 11 | 3% |
| Changed destinations and/or purpose (e.g. changed job, retired, shop locally, moved to a new house ect.) | 9 | 3% |

*Multiple Responses; Base=respondents who expects to use buses or trains less often in the future (n=333)

Considering public transport, 36% of respondents anticipated using buses less and 34% using trains less. For those who indicated that they would use public transport less, 63% considered the possibility of getting infections, carried by other passengers, would be a contributing factor; 49% considered lack of cleanliness would be a factor and 45% stated overcrowding as a factor.

Consideration was given to the factors affecting the future public transport (bus and train) travel intentions of Scottish residents. Two possible outcomes were investigated: (i) those who intend to use public transport less than they did before COVID-19, and (ii) those who intend to use public transport around the same amount, or more than they previously did before COVID-19. The research showed that several factors, including pre-lockdown travel choices, household size, age and region, influenced intended future use of public transport:

- Those who travelled by car (at least three times per week) pre-lockdown, were significantly more likely to travel less by bus and train in the future, compared to those who did not previously travel frequently by car.
- Those who travelled by a public transport mode (at least once per week) prior to lockdown, were significantly more likely to travel less by bus in the future.

The effects of influential socio-demographic factors were as follows:

- those who live in Lothian were significantly more likely to travel the same or more by bus, compared to those living in other regions of Scotland;
- young respondents (under 25) were significantly more likely than other age groups to travel more by train in the future; and
- those with a household size of three or more people were significantly more likely to travel the same or more by train in the future, in comparison to smaller households (less than three people).

### 4.4. Transport Related Lifestyle Issues

#### 4.4.1. Working from home

Respondents were asked if, during Lockdown 2, they were doing activities 'more often' or 'less often' than before the COVID-19 pandemic. As displayed in Figure 4.11, 61% of the sample worked from home more often (either 'somewhat more often' or 'much more often) and 65% indicated that they used IT technology more often in order to communicate with colleagues and/or clients.

Survey participants were asked to what extent they agreed or disagreed with statements regarding the impact of COVID-19 on their activities in the future, 12 to 18 months from now. It may be seen from Figure 4.12 that 64% of employed respondents' indicated that they will use technology to communicate with colleagues, customers or clients more and 54% of employed respondents indicated that they expect to work from home more in the future.

### 4.4.2. Online shopping

During Scotland's second lockdown, 59% of respondents are more likely to shop online to purchase products they would normally buy in-store and half (50%) were more likely to use home delivery for supermarket shopping than before the COVID-19 pandemic.

When asked about their anticipated future shopping habits, 45% indicated that they expect to do more online non-grocery shopping in the future and 36% anticipate using home delivery for supermarket shopping more.

### 4.4.3. Changes in other activities which have an impact on travel

Survey respondents indicated that 50% would go on vacation and travel less often after the COVID-19 pandemic when compared with before and only 11% would consider purchasing an additional car. However over 58% agreed with the statement "In the future, I expect to exercise outside (e.g., walking, jogging, running, cycling etc.,) more than I did before COVID-19";

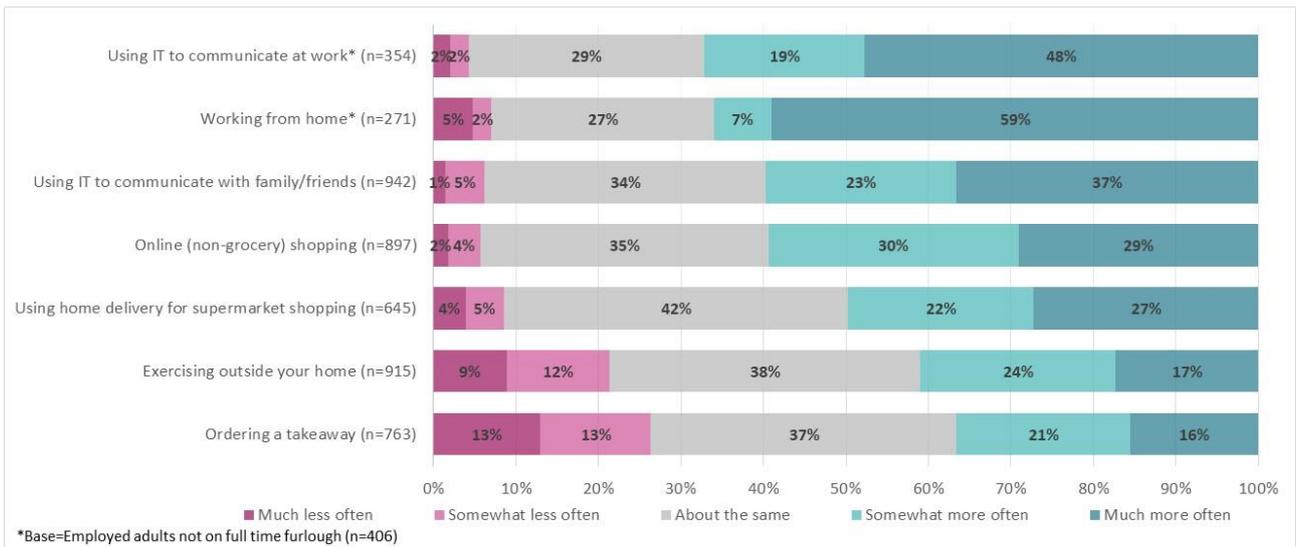

**Figure 4. 11 Activities during lockdown 2 compared with before COVID-19**

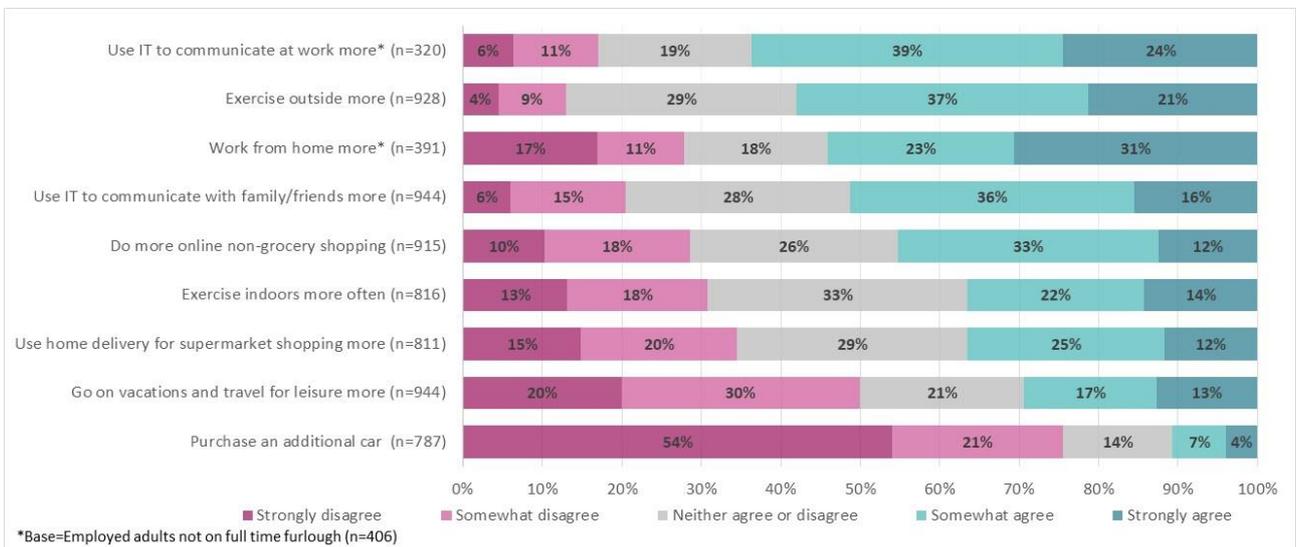

**Figure 4. 12 Anticipated future travel related activities**

## 4.5. Residential Property Relocation

In the United States of America it appears that many are in the process of moving out of city centres following the onset of COVID-19 (Gupta et al 2021). The results of the current questionnaire survey have indicated that before COVID-19, three-fifths of respondents (67%) did not expect to move from their current property in the future. When asked if they are currently (i.e., during lockdown 2) considering moving house, 60% did not anticipate ever moving to a new house.

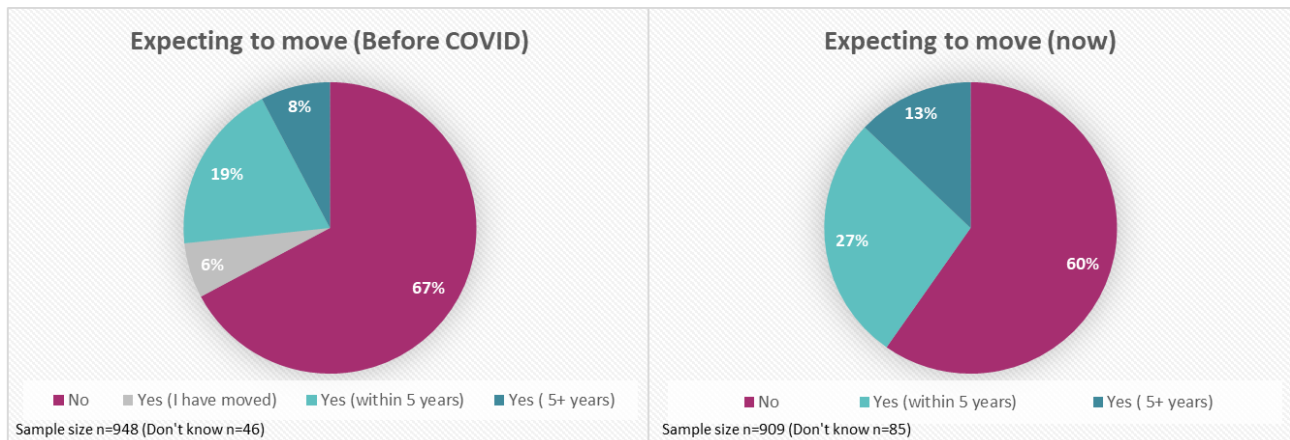

**Figure 5. 1 Proportion of respondents planning to move from current property**

Respondent who indicated that they expect to move from their current accommodation were asked where they would like to move to. The majority (57%) would like to stay in the same local area, but move to different accommodation, 33% would like to move from their local area, but stay in Scotland and 10% would like to move away from Scotland.

Table 5.1 and Table 5.2 provide information on respondents' reasons for wanting to move from their current property. The main reasons for wanting to move, in rank order, are to be near family/friends, wanting a larger property, garden/land, to buy/rent a place, move to a more rural area or move to a nicer area.

Respondents were asked which circumstances influenced them wanting to move from their current accommodation. The most frequently cited circumstances were, in rank order, wanting a different lifestyle, change in family/household circumstances, being able to work from home more often or permanently, change in employment situation and the COVID-19 pandemic.

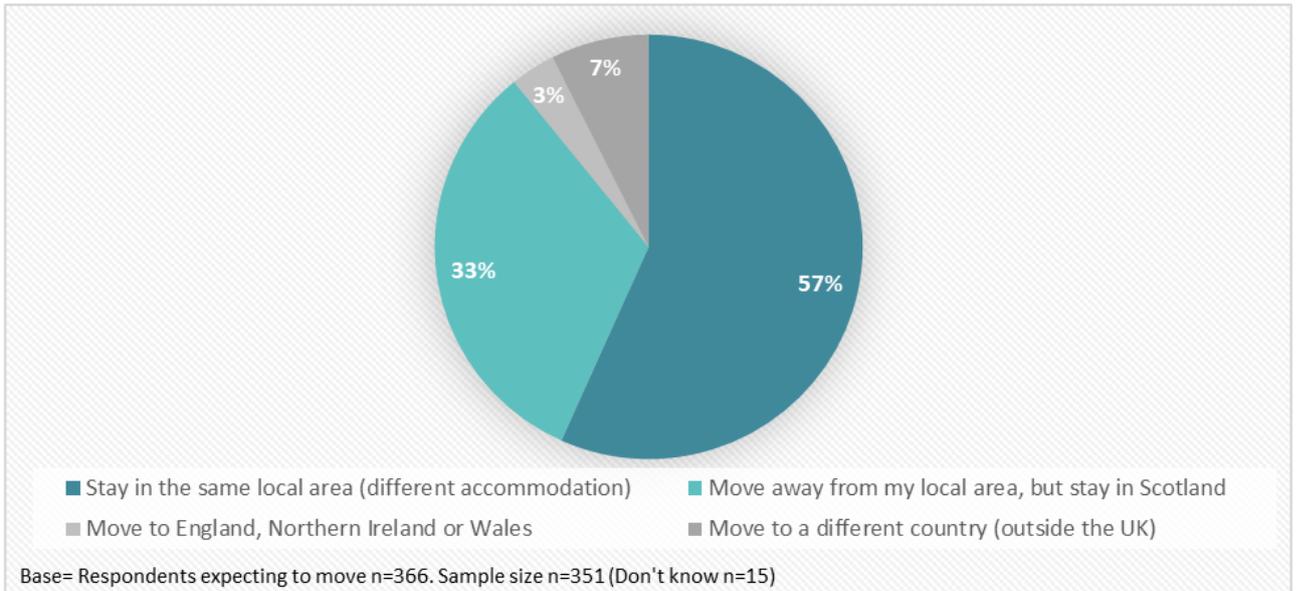

Figure 5. 2 Destination of respondents anticipating moving

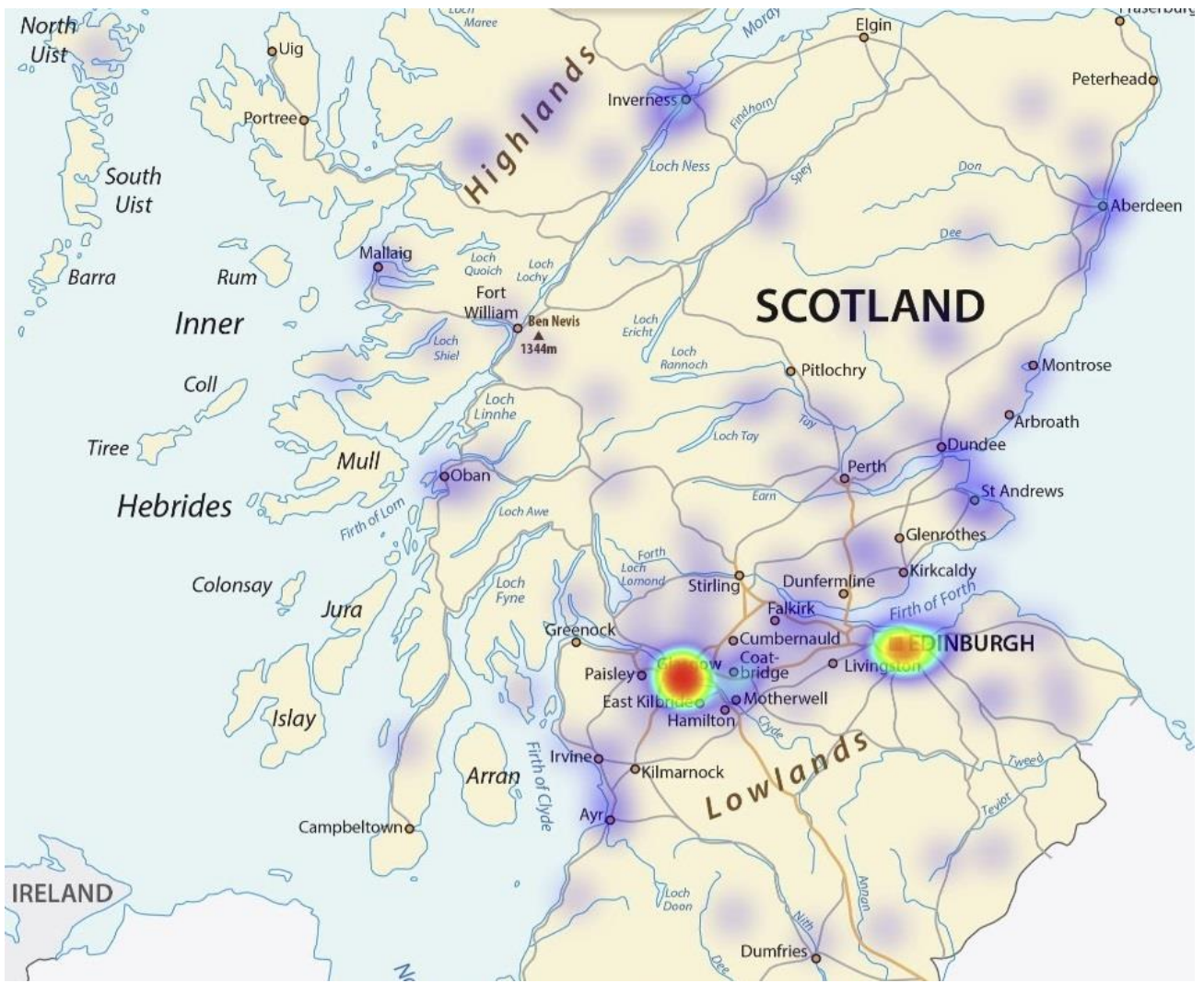

Figure 5. 3 Respondents desired destination in Scotland

**Table 5. 1 Reasons for wanting to move from current accommodation/residential location**

| Reason | Number of Responses* | Percentage (n=366) |
|---|---|---|
| To be near family/friends | 105 | 29% |
| To buy own house/flat or rent place of own | 91 | 25% |
| Want a larger property/more rooms | 82 | 22% |
| Want a garden/land | 80 | 22% |
| Move to a nicer area | 65 | 18% |
| Move to a more rural area (e.g., countryside) | 59 | 16% |
| Want a cheaper property | 53 | 14% |
| To be close to work/employment | 51 | 14% |
| To be near good services/amenities | 49 | 13% |
| Can afford a more expensive property | 44 | 12% |
| Want a smaller property | 42 | 11% |
| Move to a safer area (lower crime rate) | 34 | 9% |
| Move to a more urban area (e.g., city or larger town) | 33 | 9% |
| To be near good transport | 32 | 9% |
| Want a newer property | 29 | 8% |
| Want better parking availability | 21 | 6% |
| Move to the seaside | 18 | 5% |
| To be near good schools | 17 | 5% |
| To be close to university/college or equivalent | 12 | 3% |
| Brexit, Scottish Independence or local authority | 7 | 2% |
| Weather and/or climate | 4 | 1% |
| Other | 19 | 5% |

*Multiple Responses; Base=respondents who expect to move from their property in the future (n=366)

**Table 5. 2 Circumstances for wanting to move from current accommodation and/or residential location**

| Reason | Number of Responses* | Percentage (n=366) |
|---|---|---|
| Want a different life style (e.g., more relaxed, more active etc) | 155 | 42% |
| Change in family/household circumstances (growing family, kids leaving home etc.) | 130 | 36% |
| I can work from home more often or permanently | 62 | 17% |
| Change in employment situation (e.g., started working, changed job, retirement etc.) | 61 | 17% |
| The COVID-19 pandemic | 39 | 11% |
| Health reasons (including move to bungalow flat) | 33 | 9% |
| I will be able to use a car more often | 15 | 4% |
| Lease at current accommodation will expire | 9 | 2% |
| Move to sheltered housing/supported accommodation | 9 | 2% |
| Climb up the property ladder | 5 | 1% |
| Neighbours | 3 | 1% |
| Brexit or Scottish Independence | 3 | 1% |
| Fresh start or making a change my life | 2 | 1% |
| Other | 20 | 5% |

*Multiple Responses; Base=respondents who expect to move from their property in the future (n=366)

## 5. DISCUSSION

### 5.1. Introduction

This study examined travel behaviour, attitudes and perceptions in Scotland after COVID-19 was declared a pandemic. UK and Scottish Governments imposed 'lockdown' restrictions have resulted in unprecedented numbers of people furloughed or working from home and most out-of-home leisure activities cancelled. As a result, travel demand decreased and there has been significant reductions in both car travel and public transport use. The survey responses confirmed that there had been significant reductions in weekly car travel (69% to 55%), bus use (27% to 9%) and train use (12% to 4%) during lockdown (February 2021), by the respondents, compared to pre-pandemic levels.

Lockdown is only a temporary situation, and it may be assumed that travel demand will increase once restrictions are eased. However, Tirachini and Cats (2020) suggested that the economic and social consequences of the COVID-19 outbreak on public transport will extend beyond service performance and health risks to financial viability, social equity, and sustainable mobility.

An understanding of how COVID-19 might change travel choices in the near future and beyond will assist the Government and transport providers wishing to enhance the resilience and efficiency of the transport systems. This study provides information to assist in developing appropriate COVID-19 transport related responses, including policies, interventions and communications in the short term (e.g., advising the government on channels of communication, advising public transportation industry on protective measures) and long term (e.g. residential relocation, working from home, expected demand for public transport).

### 5.2. Life Satisfaction

It might be anticipated that the changes brought about by responses to the onset of the COVID-19 would have an impact on mental health and the feeling of wellbeing. Routen et al (2020) reported that, in the UK context, mental health declined during the COVID-19 pandemic and the associated lockdown measures. In another UK study, Pierce et al (2020) found that there was an increase in mental distress during the pandemic which they also attributed to the severe restrictions on social contact, impact on employment, greatly reduced access to services, coupled with a substantial amount of worry about future insecurity.

Both Routen et al (2020) and Fancourt et al (2021) identified a decline in the feeling of wellbeing in the UK during the COVID-19 pandemic. Similarly, as reported in Section 4 of this report, two thirds of the current study's respondents had indicated that life had got worse after the onset of COVID-19 pandemic. Over a quarter (26%) felt their lives were 'about the same' and only 7% thought their life had improved. Regarding the impact of COVID-19, over two-thirds (69%) thought that COVID-19 had changed their lives completely.

## 5.3. Communications and Perceptions of Risk

### 5.3.1. Introduction

Perception of risk affects the level of compliance with pandemic regulations and guidance and propensity to travel and is dependent on good communications. In an Australian study, associated with an influenza pandemic, Barr et al (2008) found that those with higher levels of threat perception were significantly more likely to be willing to comply with specific public health requirements. Tan et al found that the during the COVID-19 pandemic, the possibility of infection was a significant influence on commuter mode choice. Brug et al (2009) noted that effective management of risk associated with new epidemic infectious disease, when no treatment or vaccination is initially available, depends upon precautionary behaviour of the population and that this precautionary behaviour is largely dependent on effective risk communication.

### 5.3.2. Communications

In terms of government risk communications during a pandemic, Lee (2009) reported that during the 2003 SARS pandemic the public trust in Hong Kong Government communications was lost by its failures in demonstrating its effectiveness in both prevention and containment, the lack of strong leadership and it's apparent lack of commitment to the public good. There were also failures associated with the need for flexibility and good communication between government agencies and the minimising of the effects of the external crisis environment on government decisions.

As reported in Section 4, in the current study 86% of respondents considered they were well informed about the COVID-19 pandemic. In terms of information channels, 70% of the respondents 'often' or 'always' used tradition broadcasters (radio and television) for information. In contrast, only 27% regularly used newspapers as a source of COVID-19 information. Respondents had a high level of trust in medical experts (80% of respondents selected either 'extremely' or 'very' trustworthy). However, trust was much lower in communications from government sources with 51% of respondents considering the Scottish Government to be a very trustworthy source of information and 26% of respondents considering the UK Government to be a very trustworthy source.

### 5.3.3. Perceptions of Risk

In Japan, Parady et al (2020) found that concerns regarding the perceived risk of Covid-19 are associated with increases in the probability of staying at home and the probability to reduce trip frequencies. They noted shopping frequency was not reduced despite the perceived risk because such visits are essential. In such circumstance, individuals might deploy mitigating measures such as wearing masks and maintaining social distancing.

Neubuger and Egger (2020) investigated travel risk perception and travel behaviour in Germany, Austria and Switzerland during the COVID-19 pandemic in 2020. As part of the study, they found that, at a time towards the end of March 2020, after a pandemic had been declared and when the impact of COVID-19 in Europe was severe, 65% of the population could be categorized as 'anxious', showing the highest risk perception of COVID-19, travel risk perception and the highest intentions to change or cancel travel plans. The 'nervous' accounted for 21% of the population. They also had high travel risk perceptions and high intentions to change travel plans. However, they did not perceive the risk of COVID-19 was as high as those in the 'anxious' group. The 'relaxed' accounted for 22% of the population and had the lowest risk perceptions of COVID-19, lowest travel risk perceptions and lowest intentions to change or cancel travel plans.

In the United States, Hotle et al (2020) found that individuals proactive with their health (i.e., receive the vaccine, have health insurance) are also proactive in seeking medical attention and reducing influenza spread. It was also found that individuals reduce travel to locations in which they perceived a medium or high-risk existed. However, increased risk, perceived at work locations, did not significantly reduce travel. Tan et al (2020) found that, during the COVID-19 pandemic, the possibility of being infected in private cars and the possibility of being infected in public transport have significant influence on the commuters' choice of rail transit. Shamshiripour et al (2020) found that 78% of those surveyed considered travelling by bus, train or tram, during the Covid-19 pandemic, to be either a high-risk or an extremely high-risk act. Percentages for perceptions of high or extremely high-risk, for other transport modes, included shared taxis (79%), shared scooters and mopeds (52%), cycling (47%), walking (33%) and travelling in own vehicle (6%).

As reported earlier, the current survey results found that 64% would describe COVID-19 as 'fear inducing' and almost half (48%) felt COVID-19 would be a danger to them. Most respondents viewed walking (92%) and cycling (93%) as low risk. However, with reference to travelling by bus or train, 64% and 57% or respondents respectively considered these to be high-risk activities. Travelling by aeroplane was considered the most dangerous mode with 84% of respondents considering it a high-risk activity.

## 5.4. Compliance with COVID-19 Regulations and Guidance

During a pandemic, when only few if any effective treatments or vaccinations are available, its impact has to be controlled by changing a population's behaviour. As a consequence, in the COVID-19 pandemic, the Scottish Government has introduced COVID-19 regulations and guidance on travel and transport to change the Scottish population's behaviour. More than four-fifths (83%) of respondents in the current study, consider they have largely complied with these regulations and guidance with only 17% indicating that have not completely complied (with a score of 5 or less). This is similar to the findings of a study conducted by Transport Scotland (January 2021) which found that 92% of participants considered that they had completely or nearly completely followed the regulations and guidance on travel and transport. There may be concerns, with self-reporting in such a study, that respondents may avoid disclosing socially undesirable behaviour. However, Larsen et al (2020) have reported that, by comparing measures of compliance from direct questions to those estimated from list-experiments, there was no evidence that respondents will under report non-compliance. In addition, the absence of an interviewer in the current study mitigates the risk of under reporting compliance.

In terms of the level of compliance with COVID-19 regulations and demographic factors, many researchers have found that there is higher compliance amongst older members of the population (Berg-Beckhoff et al 2021, Muto et al 2021, Coroiu 2020). A possible explanation of this is younger people perceive that they have lower risk of developing a severe disease or dying when infected with COVID-19 when compared with older respondents.

In the current study, a statistical model was established to investigate the relationships between compliance with Government COVID-19 regulations and demographic variables, trust in government and the use of media as an information source (see Section 1.4). As with other studies, the model suggested that significantly more of the elderly respondents (over 65 years of age) complied with COVID-19 regulations when compared with those under the age of 25 years. In terms of other demographics, the models also suggested that both females and those unable to work because long-term illness of disabilities were significantly more likely to comply with COVID-19 regulations.

Brug et al (2009) noted that the necessary changes to behaviour during a pandemic are dependent on effective risk communication. The current study suggested that those who frequently use television, radio, websites or online sources were significantly more likely, than those who don't use this information source, to comply fully.

In Norway, Carlsen et al (2020) found that trust in the government assures people that guidelines are necessary and effective. The current research found that those who consider the Scottish Government to be a very trustworthy source, with regards to information related to COVID-19, were significantly more likely to comply fully when compared with those who perceive the Scottish Government as untrustworthy.

### 5.5. Modal Choice

#### 5.5.1. Mode choice during COVID-19 restrictions

Tirachini and Cats (2020) suggest that reduced public transport use since COVID-19 has been exacerbated by the perception of public transportation as riskier than private or personal means of transport because of the closer contact to other people that is possible, sometimes unavoidable, in public transportation vehicles and stations. In the current study it was found that, since the onset of COVID-19 there was decrease in those walking (at least half a mile) at least once a week from 81% to 75%. The percentages remained the same for cycling and those driving their own cars reduced from 69% to 55%. For those using public transport at least once a week, the number using buses fell from 27% to 9% and the percentages using trains fell from 12% to 4%. During lockdown 2 (February 2021) 43% of respondents reported using buses less often when compared with lockdown 1. The same percentage (43%) reported using trains less often. This perhaps reflected the perceived safety concerns associated with public transport. Abdullah et al (2020) have found that specific respondent variables such as gender, car ownership, employment and travel distance were significant predictors of mode choice during the COVID-19 pandemic.

### 5.5.2. Post COVID-19 Modal Choice

Responses in the current study suggested that there is the potential for a modal shift away from public transport and an increase in car dependency. A quarter of survey respondents expected to drive their car more post COVID-19 (i.e., in 12 to 18 months). Over a third of survey respondents indicated that they intend to use public transport less (36% using buses less and 34% using trains less) in the post-COVID future.

Active travel, which seeks to promote healthy journeys, offers viable alternatives to the private car or public transport for short journeys and an opportunity for exercise. Almost half (45%) of survey respondents anticipate walking more often in the future and 29% expect to cycle more often.

For those who indicated that they would use public transport less, 63% considered the possibility of getting infections carried by other passengers would be a contributing factor; 49% considered lack of cleanliness would be a factor and 45% stated overcrowding as a reason. The current study suggested several factors influenced intended future use of public transport including pre-lockdown travel choices, household size, age and region.

To reduce the spread of COVID-19 on public transport a number of mitigation measures have been proposed. In the current study, the preventative measures perceived by respondents to be most important were hand hygiene (93% rated either 'very important' or 'extremely important') and cleaning of station stops and vehicles (93%).

Gkiotsalitis and Cats (2020) highlight the importance of cleaning of public transportation vehicles in light of medical research on viruses remaining on surfaces and the transmission of micro-organisms. The current research found that those who indicated that they would use public transport less, 63% considered the possibility of getting infections carried by other passengers would be a contributing factor and 49% considered lack of cleanliness would be a factor.

The UK government instructs the public to maintain a minimum distance of 1-2 meters to reduce the chance of virus transmissions. Gkiotsalitis and Cats (2020) argue that of all the measures introduced, this is the most consequential measure for public transport, with operators having to transform their services in order to adhere to physical distancing requirements. The current research confirms that not adhering to social distancing is perceived by respondents as a potential barrier to future public transport use. Social distancing between people at stops (92%) and on-board public transport (93%) were viewed as 'very important' or 'extremely important' by respondents. Out of those who indicated that they would use public transport less, almost half (45%) stated overcrowding as a reason.

## 5.6. Transport Related Lifestyle Issues

### 5.6.1. Working from home

The emergence of the broadband internet over the last 20 years has made working from home a practical proposition. In the UK, Chung et al (2020) found that approximately a third of employees surveyed were able to work from home, at least on some occasions, before the COVID-19 lockdown. They found that the perceived benefits of working from home included improved productivity, the ability to take care of children, reduced commuter time and improved wellbeing. However, many of those surveyed in their study had problems with unsuitable working environments at home and missed interactions with colleagues.

During the COVID-19 pandemic, both governments and the private sector have encouraged working from home (Bouziri 2020). In the current study, over three-fifths (61%) of respondents indicated that were working from home more often than they were before the pandemic. Similarly, Beck (2020), Shamshiripour et al (2020) and de Hass et al (2020) found that the percentage of the workforce working entirely from home had increased during the pandemic.

De Hass et al (2020) found that the drastic shock to daily life may permanently alter the way people work and travel with many workers expecting to work more frequently at home and to continue holding remote meetings in the future. Bonacini et al (2021) have predicted that a post pandemic shift to working from home may result in an increase in average earnings with older, higher educated males benefitting most. Chung et al (2020) suggested that those from households without children were more likely to work from home in the future compared with those with children. In the current study, 54% of respondents indicated that they expected to work from home more in the future.

### 5.6.2. Shopping online

With reference to online grocery and food orders, Shamshiripour et al (2020) reports on large increases, in the USA, during the COVID-29 pandemic. Before the pandemic only 7% of respondents mostly or almost always had their groceries delivered following an online order. During the pandemic this increased to 29%. Large increases in on-line supermarket shopping involving home delivery, in Scotland, were also reported by the respondents in the current study. Half (50%) of the respondents indicated, that during the second lockdown, they were more likely to use home delivery for supermarket shopping when compared with pre COVID-19 pandemic conditions. Furthermore, when asked about their future shopping habits, 36% anticipated using home delivery for supermarket shopping more.

## 5.7. Relocating Residences After COVID-19

Allen-Coghlan et al (2020) have speculated that the potential increase in the number of people who can and will work from home in the future may have significant implications for the housing market and the general economy over the longer term. Gupta et al (2021) found that for US cities, the COVID-19 pandemic led many residents to flee city centres to the suburbs in search for safer ground away from urban density. Such shifts will have the potential to have significant impacts on overall traffic activity. As indicated earlier, in current study 33% of respondents indicated they, were prior to COVID-19 pandemic, contemplating moving from their current residence, in the future. During the COVID-19 pandemic this had increased to 40% with 17% of respondents who expect to move from their property in the future citing 'being able to work from home more often or permanently' as a contributing factor.

## 5.8. Conclusions

The COVID-19 pandemic has brought considerable changes to both the nature of travel behaviour and transport in Scotland. The present research set out to investigate travel choices, attitudes and perceptions and transport related lifestyle issues (e.g., online shopping, working from home, moving to a new house) during and after the pandemic.

The main findings indicated that the level of cycling remained constant, for the respondents during the pandemic with slight reductions in walking and driving a car. In contrast, there were very large reductions in the use of buses, trains and aeroplanes. After the pandemic, 45% of respondents expected to walk more than before the pandemic and 25% of respondents expected to drive their car more. However, 42% of respondents anticipated using aeroplanes less, 36% anticipated using buses less and 34% anticipated using trains less.

The investigation also found many respondents reported a fall in their 'life satisfaction' following the onset of COVID-19. There was also a large increase in the number of respondents working from home, the number of respondents using online shopping for groceries and the number of respondents expecting to move from their current properties. The respondents also indicated that they anticipated that they would work from home more and use online supermarket shopping facilities more after the pandemic when compared with before.